\documentstyle[11pt,newpasp,twoside,psfig,epsf]{article}
\makeindex
\index{a:Braine, J.}
\index{a:Duc, P.--A.}
\index{a:Lisenfeld, U.}
\index{a:Brinks, E.}
\index{a:Charmandaris, V.}
\index{a:Leon, S.}
 
\index{k:Galaxies-general!dark matter}
\index{k:Galaxies-general!dwarf}
\index{k:Galaxies-activity!star formation rate}
\index{k:Tidal Dwarf Galaxies!ISM}
\index{k:Tidal Dwarf Galaxies!dark matter}
\index{k:Galaxies-general!evolution}
\def\edcomment#1{\iffalse\marginpar{\raggedright\sl#1\/}\else\relax\fi}
\reversemarginpar

\begin{document}
\title{Molecular Gas in Tidal Dwarf Galaxies: On-going Galaxy Formation}
 \author{J.~Braine$^1$, P.-A. Duc$^2$, U. Lisenfeld$^3$,
 	E. Brinks$^4$, V. Charmandaris$^5$, S.~Leon$^3 $}
\affil{$^1$Observatoire de Bordeaux, UMR 5804, B.P. 89, F-33270
		 Floirac, France \\
	$^2$CEA/DSM/DAPNIA, Service d'Astrophysique, Saclay, France\\
	$^3$IAA, CSIC, Granada, Spain \\
	$^4$INAOE, Puebla, Mexico \\
	$^5$Cornell University, Astronomy Department, Ithaca, NY, USA
}

\begin{abstract}
We investigate the process of galaxy formation as can be observed in the only
currently forming galaxies -- the so-called Tidal Dwarf Galaxies, hereafter
TDGs -- through observations of the molecular gas detected via its CO 
(Carbon Monoxide) emission. 
%These objects are formed of material torn off of the outer parts of a spiral 
%disk due to tidal forces in a collision between two massive galaxies.
Molecular gas is a key element in the galaxy formation process, 
providing the link between a cloud of gas and a {\it bona fide} galaxy.  
We have now detected CO in 9 TDGs with an overall detection rate 
of 80\%, showing that molecular gas is abundant in TDGs, up to a few 
$10^8 M_\odot$.  The CO emission coincides both spatially and 
kinematically with the HI emission, indicating that the molecular gas 
forms from the atomic hydrogen where the HI column density is high.
A possible trend of more evolved TDGs having greater molecular gas
masses is observed, in accord with the transformation of HI into H$_2$.
%
%$\, \, \, \, \, \, $
%Although 
%TDGs share many of the properties of small irregulars, their CO 
%luminosity is much greater (factor $\sim 100$) than that of standard
%dwarf galaxies of comparable luminosity. This is most likely a 
%consequence of the higher metallicity ($\ga$ 1/3 solar) of TDGs which
%makes CO a good tracer of molecular gas. This allows us to study star
%formation in environments ordinarily inaccessible due to the extreme 
%difficulty of measuring the molecular gas mass. The star formation
%efficiency,  
%measured by the CO luminosity per H$\alpha$ flux, is the same in TDGs
%and full-sized spirals. \\
%
%$\, \, \, \, \, \, $
%CO is best tracer of the dynamics of these objects
%because some fraction of the HI near the TDGs may be part of the tidal 
%tail and not bound to the TDG.  
Although uncertainties are still large for individual objects as the
geometry is unknown,
%, our sample is now of nine detected objects and 
we find that
the ``dynamical" masses of TDGs, estimated from the CO line widths, 
do not seem to be greater than the ``visible" 
masses (HI + H$_2$ + a stellar component),  
%Although higher spatial
%resolution 
%CO (and HI) observations would help reduce the uncertainties, we find that
i.e., TDGs require no dark matter.  
%Dark matter in spirals 
%should then be in a halo and not a rotating disk.  \\
%Most dwarf galaxies 
%are dark matter-rich, implying that they are {\it not} of tidal origin.
%
%$\, \, \, \, \, \, $
We provide  evidence that TDGs are self-gravitating entities, implying 
that we are witnessing the ensemble of processes
in galaxy formation: concentration of large amounts of gas in a bound object,
condensation of the gas, which is atomic at this point, to form molecular
gas and the subsequent star formation from the dense molecular component.
\end{abstract}

\section{Introduction} 

Tidal Dwarf Galaxies (TDGs) are small galaxies which are currently forming
from material ejected from the disks of spiral galaxies through collisions.
They allow us to observe processes -- galaxy formation and evolution -- similar
to what occurred in the very early universe but in very local objects.  As a
consequence, they can be studied with a sensitivity and a resolution 
inconceivable for high-redshift sources.  Because galactic collisions 
can be well reproduced through numerical simulations, it is possible 
to obtain good age estimates for the individual systems (e.g. Duc et al. 2000).  
The formation of TDGs is not exactly the same as what happened during 
the major episode of galaxy formation in that the material
which TDGs are made from is ``recycled", as it was already part of a
galaxy.  In particular, the presence of metals, both in gas and as dust, 
facilitates the cooling of the gas and the formation of H$_2$ molecules.
Nevertheless, both for TDGs and in the early universe, the galaxy 
formation process involves clouds of atomic hydrogen (HI)
gas gradually condensing through their own gravity, becoming progressively
denser, fragmenting, forming molecular gas from the atomic material, and
then forming stars.  How this occurs in detail at high redshift is 
unknown and is one reason for studying TDGs.

Perhaps the least well known of these processes is the transformation of atomic
into molecular gas because of the difficulty of observing molecular gas in 
very low-metallicity environments (e.g. Taylor et al. 1998).  
Because TDGs condense from matter taken from the outer disks of spiral galaxies, 
the metallicity of the gas they contain is typically only slightly subsolar 
as opposed to highly subsolar for small dwarf galaxies (Duc et al. 2000).  
The metallicity dependent CO lines can thus be used as a probe
of the molecular gas content as in spiral galaxies.
The study of TDGs influences three areas of astronomy: star formation, dark 
matter, and galaxy formation.

Much of the material described here has been published in Braine et al. 
(2000, 2001; hereafter Papers I and II.).
Readers are referred to these articles for the lengthy details of the sample
as well as many figures showing the CO spectra.  Here we focus on the results
and the most recent detection.  

\begin{figure}[h]
\psfig{figure=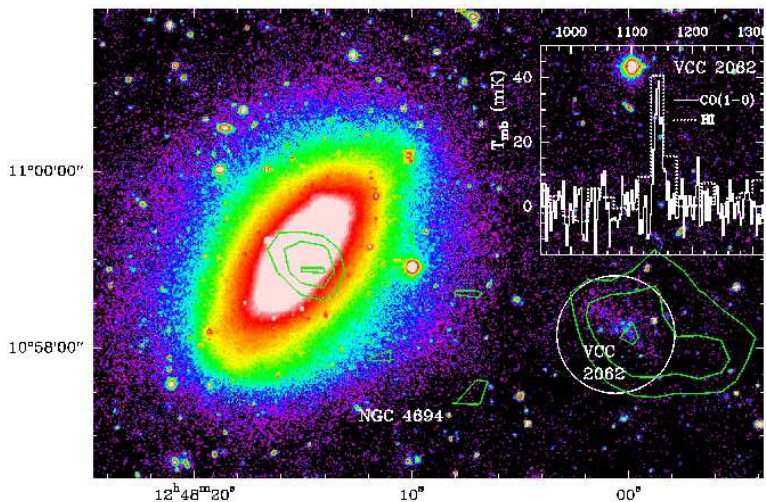,width=120mm,angle=270.0}
\caption{R band image of NGC 4694 and VCC 2062 from Koopman et al. (2001) 
with the CO(1--0) and HI spectra of VCC 2062 inset.  The angular resolutions are
21\arcsec\ and 38\arcsec\ respectively.}
\vspace*{-0.5cm}
\end{figure}

\section{Results}
\index{o:VCC 2062}
\index{o:NGC 4694}
\index{o:Stephan's Quintet (HCG 92)}

Our sample consists of interacting systems for which an extensive set of
optical and radio (HI) data already exist in the literature. 
CO detections were obtained for the following TDGs: VCC 2062 (west of
NGC~4694 in the  
Virgo cluster), NGC~7252 West (hereafter NGC7252W), Arp105S, Arp245N,
NGC~4676 North (hereafter NGC4676N), NGC~5291 North and NGC~5291 South
(hereafter NGC5291N and NGC5291S), Stephan's Quintet source 'B' (hereafter 
NGC7319E), and very probably NGC~4038/9 South (``The Antennae", 
hereafter NGC4038S).  Figure 
1 shows an R band image of the VCC~2062/NGC~4694 system with contours 
showing the HI emission from Cayatte et al. (1990). An inset shows the
CO and HI spectra at the position of VCC~2062.  The other systems are 
shown in Papers I and II.
%All coordinates are given in the J2000 coordinate system.
The western tail of the NGC~2782 (Arp~215) system was also 
observed with no CO detection, confirming the 
Smith et al. (1999) non-detection.  The western HI tail of NGC~2782 has 
presumably not had time to condense into H$_2$ and for star formation to 
begin (see below).  
We use a $N({\rm H}_2) / I_{\rm CO}$ factor of 
$2 \times 10^{20}$cm$^{-2}({\rm K} {\rm \, km \, s^{-1}})^{-1}$
for all calculations of H$_2$ masses from CO measurements.

\section{Star Formation in TDGs and other Dwarf Galaxies}

The most striking difference between TDGs and Dwarf Galaxies not identified as 
tidal is their high CO luminosity (see Fig. 2a), roughly a factor 100
higher than for other dwarf galaxies of similar luminosity and star 
formation rate (SFR).  The most important factor responsible 
for this difference is certainly the metallicity.  As the metallicity
increases, the CO lines become optically thick over a larger area and 
larger velocity range.  Furthermore, the shielding against UV radiation due 
to both CO molecules and dust increases as well.  For this reason, 
a change in metallicity is expected to have a stronger effect in a 
UV-bright environment (e.g. Wolfire et al. 1993; Braine et al. 1997 Sect 7.3). 
The details of the comparison dwarf galaxy sample and the SFRs are given 
in Paper II.

%The metallicities of
%TDGs are (from [OIII]/H$\beta$ line ratios) clustered around 
%$12+\log(O/H) = 8.5$
%independent of luminosity (Duc et al. 2000) whereas non-TDGs obey a 
%luminosity-metallicity relation (Skillman et al. 1989).  
%Inspection of the available CO data on dwarf galaxies reveals that up to now, 
%there are probably no real 
%CO detections at $12+\log(O/H) < 8$, corresponding to M$_B \ga -15$.
%The detection of several TDGs at M$_{\rm B} > -15$ 
%confirms that metallicity is indeed a key element and that 
%luminosity is not a problem for the detection of molecular gas in TDGs
%as long as sufficient HI is present.

We illustrate some of the differences between TDGs, standard dwarf galaxies, 
and normal spirals in Fig. 2a, where we show the molecular gas content 
derived from the CO luminosity, normalized by the star formation rate (from 
H$\alpha$ flux), as a function of luminosity.  Whereas the luminosity  range 
of TDGs is indeed typical of dwarf galaxies, their M$_{\rm mol}$/SFR ratio 
(equivalent to CO/H$\alpha$) is rather typical of spirals
and much higher (about a factor of 100) than in dwarf galaxies.

%In TDGs, as in spiral galaxies, CO is likely a good tracer of H$_2$. 
% and the $N({\rm H}_2) / I_{\rm CO}$ conversion factor does 
%not seem to be radically different in TDGs and in spirals. 
%The similar gas consumption times suggest that star formation proceeds in a
%similar way in both spiral disks and small irregular systems as TDGs.
It is remarkable that the gas consumption time, the inverse of 
the SFE, appears not very different in spiral disks, dominated by the 
stellar mass, and TDGs which are frequently dominated by the gaseous mass
and do not share the rotating disk dynamics of spirals.

\begin{figure}
%\psfig{figure=braine_s217_fig2.ps,width=90mm,angle=270.0}
%\plottwo{braine1.fig2.ps}{braine1.fig3.ps}
\plotfiddle{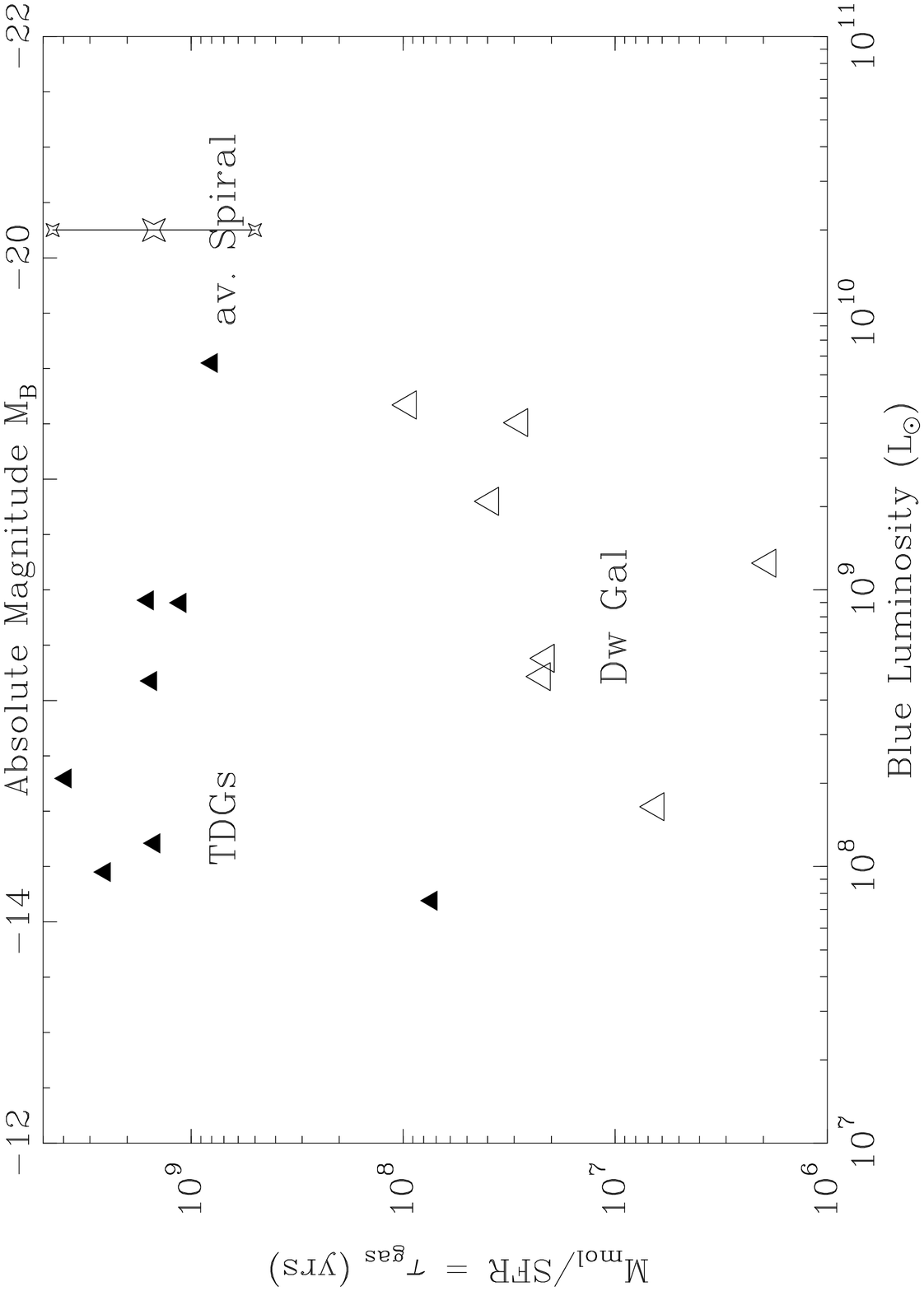}{0pt}{-90}{25}{25}{-190}{10}
\plotfiddle{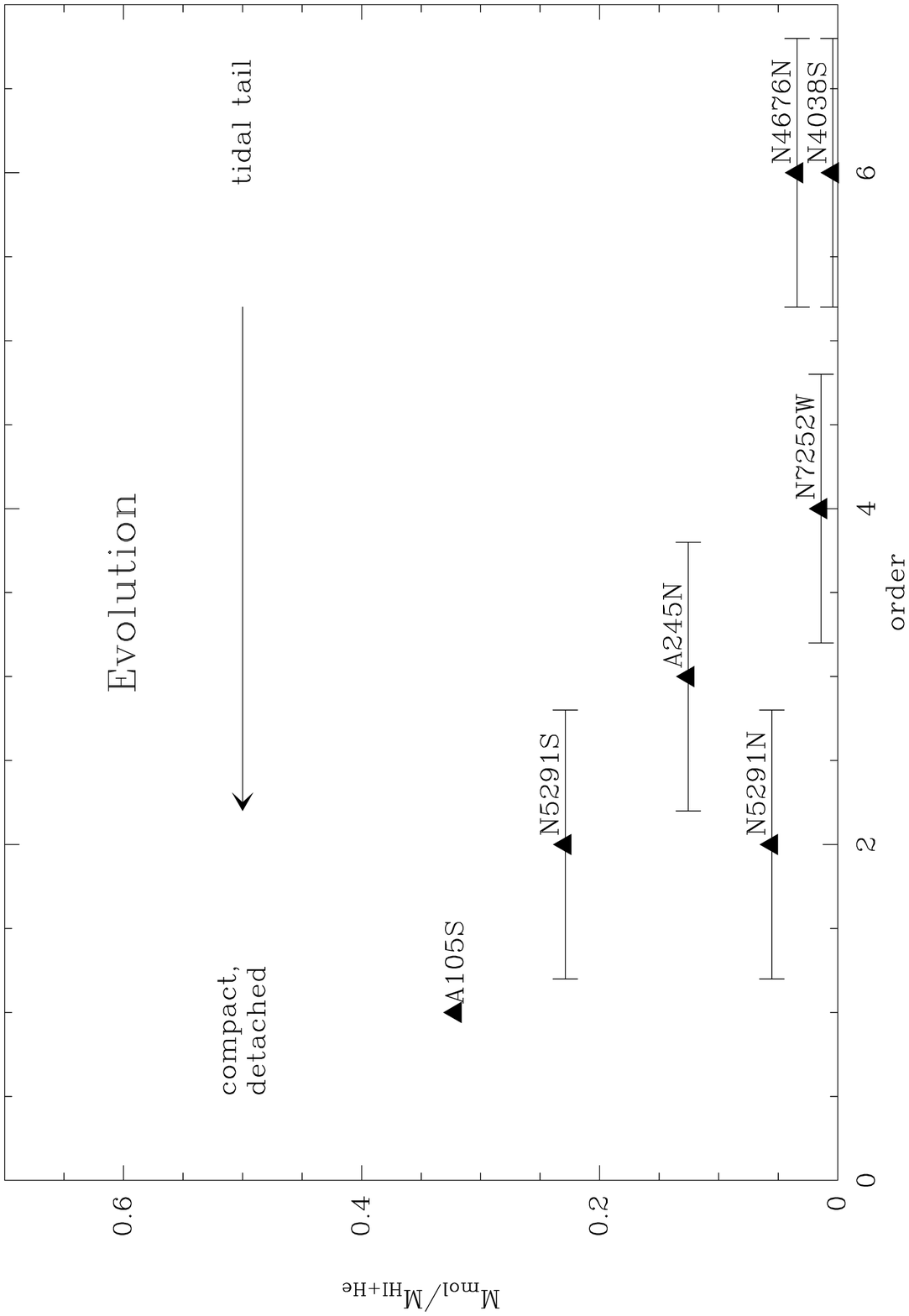}{0pt}{-90}{25}{25}{0}{30}
\vspace{100pt} 
\protect\caption{{\it left:} Comparison of gas consumption time, M$_{\rm mol}$/SFR 
(the inverse of the Star Formation Efficiency (SFE)) as a 
function of blue luminosity.  Small filled triangles represent the TDGs, 
open triangles the regular dwarf galaxies, and the star 
an average spiral, similar in position on the diagram to the Milky Way.  
The smaller stars give the range for ``average" spirals.
{\it right:} Comparison of the H$_2$/HI mass ratio with the evolutionary
order of the TDGs. NGC7319E was not included due to its unclear morphology.
The single non-detection, IC1182E, was also left out. 
The errorbars indicate the uncertainty in the evolutionary order. }
\end{figure}

%\begin{figure}
%\psfig{figure=braine_s217_fig3.ps,width=90mm,angle=270.0}
%\caption { Comparison of the H$_2$/HI mass ratio with the evolutionary
%order of the TDGs. NGC7319E was not 
%included due to its unclear morphology (see Fig. 6).
%The single non-detection, IC1182E, was also left out. 
%The errorbars indicate the uncertainty in the evolutionary order.}
%\end{figure}

\section{TDG Formation and the transformation of HI into H$_2$}

We argue here that it is now possible to follow the TDG formation
process from ejection to gravitational collapse to the conversion 
of HI into H$_2$ and the subsequent star formation.

In Paper I, presenting the first CO detections in the TDGs Arp105S and Arp245N, 
we ascribed the CO emission to the formation of molecular gas from the HI.  
We describe this in more detail below; the calculation of the H$_2$
formation time can be found in Paper II.
The time to transform 20\% of the atomic hydrogen into H$_2$
is $t_{20} \approx {{10^7} \over {n_{\rm HI}}}$ years.  
$t_{20}$ is an appropriate indicator because most of the hydrogen gas is still
in atomic form, with 20\% being a typical H$_2$ fraction.  
The HI will become molecular in the densest parts, staying atomic in less 
dense regions.
The timescale for the HI $\rightarrow$ H$_2$ conversion is thus much shorter 
than the other galaxy interaction and formation timescales, of order Myr
for typical densities (after some contraction) of $n_{\rm H} \ga 10$ cm$^{-3}$.  
The HI $\rightarrow$ H$_2$ conversion is thus not able to slow down 
the gravitational collapse.

 The average HI column density is probably not very relevant 
in and of itself.  Rather, once 
the surrounding material is gravitationally contracting, 
the HI clouds come closer together 
and provoke the transformation of HI into H$_2$, which is what allows star 
formation to proceed.  This is what we see in the TDGs.
 An interesting counterexample is the western tidal arm (or tail) of
 NGC~2782.  It clearly stems from NGC~2782 and is very HI-rich with many
 condensations with average column densities of N$_{\rm HI} \sim 10^{21}$
 cm$^{-2}$ over regions several kpc in size.
% The HI is accompanied by a weak cospatial stellar plume of surface 
% brightness about $\mu_{\rm B}\sim 25$ or perhaps slightly weaker.  
Neither we nor Smith et al. (1999) detected CO despite the high HI column 
densities and the presence of disk stars.
 
 NGC 2782 has no single big (TDG-sized) HI condensation 
 at the end of the western tidal tail and indeed the lack of CO provides
 a coherent picture: the HI here is not condensing because the tail is
 not gravitationally bound and thus H$_2$ is not forming so star formation
 has not started.  In fact, the interaction 
 has certainly added some energy to the tail so we expect that the clouds
 may separate further.  In the cases where CO is detected, the large (TDG) 
 scale is gravitationally bound and even though large amounts of H$_2$ 
 do not form in the
 outer disks of spirals, H$_2$ forms here because the HI clouds become closer
 to each other with time, pushing them to form H$_2$.
 Along with the almost complete lack of CO in the outer disks of spirals, 
 the western tail of NGC 2782 is straightforward observational 
evidence that the CO is {\it not} brought out of spiral disks.

\section{Evolution and morphology}

Assuming TDGs are not short-lived objects, they condense from the tidal tail,
the unbound parts of which slowly separate from the TDG.  TDGs can then
be arranged in a morphological evolutionary 
sequence which follows ($a$) their degree of
detachment from the tidal tail, which can be roughly defined as the density
enhancement with respect to the tidal tail, and $(b$) the compactness of
the object, which is a measure of the degree of condensation of the gas
(stars are non-dissipative so the old stellar population, if present, 
will not ``condense").  The classification (evolved, intermediate, young)
is simple for a number of objects.  Arp105S is clearly very compact and,
at the opposite end, NGC4038S and NGC4676N are only just condensing from
the tidal tail, being non-compact and with only a small density (light or HI)
enhancement with respect to the tail.  
%NGC7252W is clearly more enhanced
%and compact than either NGC4038S or NGC4676N but nothing like Arp105S.
%The same is true for the much more massive Arp245N.  For the NGC5291 TDGs
%the stellar enhancement is total and the HI enhanced by a factor of a few,
%although it is still extended; they are clearly more evolved by these 
%criteria than Arp245N.
%NGC7319E is not currently classifiable because we do not know whether 
%it belongs to an extended optical structure or not.  
The sequence suggested in Fig. 2b represents this evolution.
%, not necessarily age; simulations show, for example, 
%that the NGC~7252 merger is much older than the Arp~245 interaction 
%(Hibbard \& Mihos 1995; Duc et al. 2000).  

The conversion of HI into H$_2$ during contraction suggests that
the H$_2$ to HI mass ratio may also be a tracer of evolutionary state.  
While a real starburst may blow the gas out of a small galaxy, the H$\alpha$
luminosities do not suggest that this is the case for the TDGs here.
In Fig. 2b we plot the molecular-to-atomic gas mass ratio as a function 
of class, where increasing class indicates less evolved objects.  
%Within the criteria defined in the preceding paragraph, objects can 
%be moved around somewhat but the main subjectivity is where the NGC5291
%TDGs are placed between Arp105S and Arp245N, showing in all cases that
%the two evolutionary tracers behave in a similar fashion.  
That no known
objects occupy the upper right part of the figure is further evidence 
that the H$_2$ (or CO) does not come from pre-existing clouds in spiral 
disks but rather formed in a contracting object.
New high-resolution VLA HI and Fabry-Perot H$\alpha$ data should allow 
subtraction of the tail contribution and enable dynamical, and not
purely morphological, criteria to be taken into account (see contributions by
Amram and Duc).  We will then be able to check and quantify the 
qualitative evolutionary sequence proposed here.

\section {Dark Matter, CO linewidths and Tidal Dwarf Galaxy masses}

While data are still sparse, we believe there is a correlation between CO line 
widths and mass indicators.  Because CO is found in the 
condensed parts of TDGs, it is a better mass indicator than the HI 
linewidth, for which the contribution of the tidal tails or other unbound 
material (not TDG) cannot be easily assessed.  
Clearly, a regular HI (or CO) rotation curve would be 
extremely useful and convincing as a measure of mass.  
%So far, regular
%rotation curves have not been observed in TDGs although velocity gradients, 
%possibly rotation, in the ionized gas have been detected (Duc \& Mirabel 1998).

In order to trace the mass of a system, the material used as a tracer must
be ($a$) gravitationally bound and ($b$) roughly as extended (or more) as
the mass distribution.  We believe that CO fulfills these conditions for 
TDGs.  The fact that the CO is found where the HI column density is high,
and that it formed from the condensation of the HI, is good evidence for 
condition $a$.  The second condition is more problematic given the large
distance and small angular size of most of our sources but nonetheless 
several considerations lead us to think it is justified.  The only extended 
TDGs, with respect to the resolution of our observations, are NGC4038S and 
Arp245N.  Arp245N is roughly as extended in CO as at other wavelengths; NGC4038S 
was only observed at one position.  No abundance gradient has been detected
or is expected in current TDGs so one may reasonably expect CO to be visible 
wherever HI has condensed into H$_2$.  Many dwarf galaxies have very extended 
HI distributions, up to several times the size of their optical extent.
In TDGs, the evidence points to relatively co-spatial dense gas and 
old stellar populations although the relative distributions in the parent disk
and collision parameters condition the mass ratio.  The diffuse, unbound, 
HI in tidal tails is unlikely to form H$_2$ so the molecular component should
yield a complete but less confused picture of the dynamics.
The old stellar population of TDGs varies greatly but is in general 
quite dim, furthering the expectation that were DM present in TDGs, we would 
see it in the gas dynamics.

The apparent, although very rough, correlation between the ``Virial masses", 
$M_{\rm vir} \approx R \Delta V^2 / G$, and the HI, H$_2$, and luminous 
masses, is shown in Fig. 3.  It is an indication that the line widths are 
indeed related to mass, analogous to the Tully-Fisher relation for spirals.  
We infer this principally from the unpopulated lower right (high mass, 
low linewidth) and upper left (low mass, high linewidth)
corners of the panels.  In turn, this implies that ($a$) the objects are 
kinematically distinct from the parent galaxies and ($b$) the linewidths can 
be used as an indicator of mass.
We say indicator of mass as opposed to measure of mass because of the great 
uncertainties, factor 2 or more, in the geometry as well as in the degree 
of relaxation of the objects.  
The uncertainty is not symmetric, however, as lines can be widened more 
easily than narrowed.
Low values of  $R \Delta V^2 / G$ are thus significant.

Hibbard \& Barnes (this volume) argue that projection effects can result in 
inappropriately high virial masses (their expression for $M_{\rm vir}$ is 
roughly equivalent
to ours) and that we (Braine et al. 2001) have overestimated the contribution 
of HI to the TDG masses.  Their first point is entirely correct and is why our {\it
low} values of $M_{\rm vir}$ are significant.  Indeed, {\it if} projection 
effects had an effect on CO, then we would find high $M_{\rm vir}$ values which 
is not the case and again shows that CO is an appropriate tracer of the dynamical 
mass.  Inspection of the CO and HI spectra in Braine et al. (2000, 2001), which have
similar angular resolutions, shows that only a minority (0 -- 50\% depending on the
object) of the HI (at the HI peak) is found at velocities outside of the CO range.
One can also see this in Fig. 1 -- at the peak, the HI is only slightly
wider than the CO such that less than half is outside the CO velocity range.
This cannot change our conclusions.  The recent simulations by Bournaud et al. 
(2003; Duc et al. this volume) confirm again that TDGs condense under their own
gravity, which is generated by material from the original spiral disks.
Amram et al. (this volume) show that projection effects can be identified
kinematically and appear not to be responsible for the appearance of
several of 
the TDGs detected in CO; the TDG candidate in which we did not detect CO, 
IC 1182E, shows the signs of a projection effect.

\begin{figure}[t!]
\psfig{figure=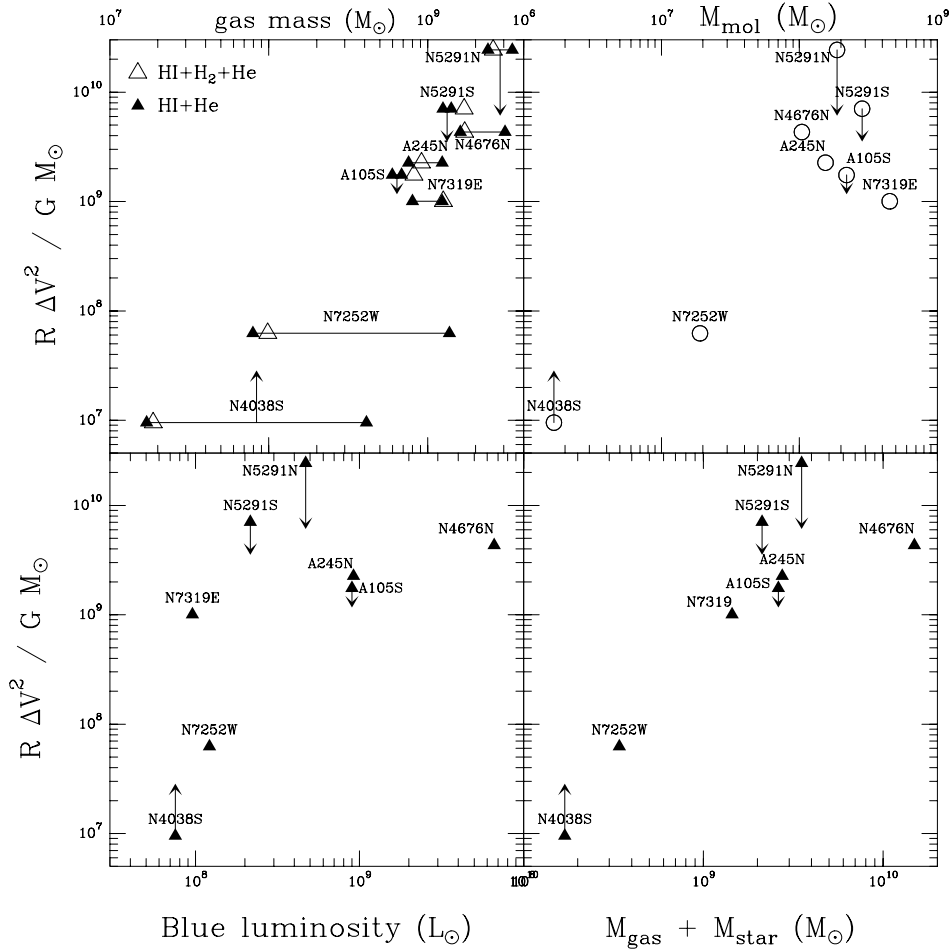,width=125mm,angle=0.0}
\caption { Virial masses derived from CO line widths as a function of HI and 
total gas mass (top left),
H$_2$ mass (top right), Blue luminosity (lower left), and a
combination of all of the above.  Ordinate axis expresses the Virial 
masses as $R \Delta V^2 / G$.
In the first panel the dark triangles joined by a line represent 
the HI masses within the area observed in CO and total HI masses.  
Open triangles give total gas masses within the CO 
beam(s) -- these are dominated by the HI.  An idea
of the uncertainty and whether the value is an overestimate or underestimate
is given by the presence of an arrow.  A downward arrow is caused by a beam which
is large compared to the source size -- such that $R_{\rm TDG} < R_{\rm beam}$
or that $\Delta V$ is likely an overestimate.  NGC4038S is 
larger than the beam but only one position was measured -- hence the up arrow.
The position of the end of the arrow is our best estimate, based on source 
size or line widths in other lines, of where the point should really
be placed in the figure.
\vspace{0.5cm}}
\end{figure}

Figure 3 shows the variation of the ``Virial mass" with total gas mass, 
molecular gas mass, Blue luminosity, and our best estimate of 
the total mass.  Tidal features
are not a homogeneous class -- some have virtually no pre-existing stellar
component (e.g. NGC5291N) while others (e.g. Arp245N) have a significant
contribution from disk stars.  To take this into account, we 
tried to sum the masses of the gaseous and stellar components in the last
panel and indeed the trend shows a smaller dispersion.  Although the mass 
to light ratio, M/L$_{\rm B}$, certainly varies within the sample we 
chose a ratio of M/L$_{\rm B} = 2 {\rm M/L_{B,\odot}}$, midway between 
that of young and evolved stellar populations.

The ``Virial masses" of the sample span a larger range than
the gas + star masses.
Some of this may be due to the uncertain line width of NGC5291N.
Given the uncertainties in line widths and 
geometry it is too early to make definite statements but so far no dark
matter is required to explain the observed CO line widths.
%$M_{\rm vir} \gg M_{\rm gas+stars}$ would imply that either the object 
%was bound by dark matter or that the object is not bound and seen
%with the projection effects described by Hibbard \& Barnes.
%Currently, our CO observations show no evidence of dark matter or
%projection effects.
It will be interesting to see whether this remains true when more objects
and more precise measurements are available.  It is now clear, however, that 
TDGs can form (condense) under their own gravity and that the uncertainties 
in our conclusions will decrease steadily as more objects are observed.

If indeed TDGs do not contain DM, then \\
\noindent --- they are the only DM-free galaxies identified so far. \\
\noindent --- DM is found in the haloes of spiral galaxies. \\
\noindent --- TDGs are not representative of the population of Dwarf Galaxies 
with measured rotation curves as these are quite DM-rich.

\end{document}